\documentclass[amsmath,amssymb,twocolumn]{revtex4}
\usepackage{dcolumn}
\usepackage{bm}
\usepackage{amsmath,mathrsfs}
\usepackage{graphicx,epsfig}
\usepackage{hyperref}
\usepackage{epsf}
\usepackage{color}

\begin{document}


\title{{\bf Wormhole in $f(Q)$ gravity }}

\author{ F. Parsaei$^1$ }\email{fparsaei@gmail.com}
\author{S. Rastgoo$^1$}\email{rastgoo@sirjantech.ac.ir}
\author{P.K. Sahoo$^2$}\email{pksahoo@hyderabad.bits-pilani.ac.in}
\affiliation{ $^{1}$Physics Department , Sirjan University of Technology, Sirjan 78137, Iran.}
\affiliation{ $^{2}$Department of Mathematics, Birla Institute of Technology and Science-Pilani,
Hyderabad Campus, Hyderabad-500078, India.}
\date{\today}


\begin{abstract}
\par  In this paper, exact asymptotically flat wormhole solutions in the context of  symmetric teleparallel gravity, i.e., $f(Q)$ theory of gravity, are investigated. Since modified theories of gravity provide new field equations, we have analyzed some possible wormhole solutions by using modified field equations. Four different forms of the $f(Q)$ function are considered then  the shape function is calculated with some different equations of state. Also we have used a power-law shape function with these models, which lead to an asymptotically barotropic equation of state. Some physical and mathematical  properties of the solutions, like energy conditions and boundary conditions, are addressed. \\
\end{abstract}
\maketitle
\section{Introduction}

Wormhole is a theoretical solution of Einstein field equations that has no horizon or singularity \cite{Visser}.  Theoretically, wormhole can connect two points in the same or different universes by speed more than the light. This fantastic property motivated many researchers to study wormholes.
There is not enough observable data which can explain the existence of the  wormholes but lensing by wormholes and light deflection have been studied \cite{lens}.
Einstein and Rosen explained the structure of the wormholes mathematically \cite{Rosen}, but  the term ‘wormhole’ has been introduced by Misner and Wheeler \cite{wheeler}. The main ingredient of the wormhole, in the context of general relativity (GR), is the violation of energy conditions \cite{WH}. The matter which violates energy conditions is called exotic.
Recent astrophysical observation indicates an accelerated expansion of the Universe which relaxed the acceptance of exotic matter. So, the phantom wormhole solutions have been considered more after this observation \cite{phantom, phantom1}. Since minimizing exotic matter is the main challenge in the wormhole theory, cut and paste method \cite{cut} or wormholes with variable equation of state (EoS) \cite{Remo,variable,foad} have been studied in the literature to minimize the exotic matter. The “volume integral quantifier” is one of the most popular approaches which quantifies the total amount of energy condition violating matter \cite{vol, zas}. On the other hand, the modified theories of gravity opened  a new window to study the wormholes. Many different modified theories of gravity have been used to investigate wormhole theory. In  most of these theories, the additional terms on the right hand side of Einstein modified field equations  provide  solutions in which the ordinary matter respect the energy conditions. Many of these modified theories of gravity have been used to explain the dark energy (DE) and, mostly, the accelerating cosmic expansion.  Since the modified theories are quite helpful in explaining the cosmic expansion and other related concepts, it will be interesting to test the ability of these theories in explaining  astrophysical objects like wormholes.  Rastall theory \cite{rast}, curvature
matter coupling \cite{curvature}, braneworld \cite{brane}, Born-Infeld theory \cite{Born}, quadratic gravity \cite{quad}, and Einstein-Cartan gravity \cite{Cartan} are some examples.
    The traversable wormholes and thin shell wormholes are studied in $f(R)$ gravity \cite{f(R)}. In $f(R)$ theory of gravity, the Ricci scalar curvature, $R$, is replaced by an arbitrary function of the scalar curvature, $f(R)$, in the gravitational Lagrangian density \cite{Rev}. Besides, the teleparallel gravity is considered to be an alternative to GR where Torsion($T$) defines the gravitational interaction. Some works are done in the context of the teleparallel gravity to study the wormholes \cite{f(T)}.
 Recently a new class of modified gravity, dubbed $f(Q)$ gravity is  proposed by Jimenez et al.\cite{Jim}. In this theory of gravity,  both curvature and torsion vanish, and gravity is attributed to non-metricity ($Q$). The fundamental difference between  teleparallel gravity and GR is the role of the affine connection, rather than the physical manifold \cite{Jim}. $f(Q)$ gravity is favored in describing the accelerated expansion of the Universe at least to the same level of
statistic precision of most renowned modified gravities \cite{Zha}.

Recently, the study on  symmetric teleparallel gravity  has been utilized rapidly in theoretical and observational fields. This modified theory of gravity has been used to explain cosmological evolutions and other aspects \cite{Tel}. The energy conditions  \cite{Ec}  and Newtonian limit \cite{Newt} in the context of $f(Q)$ are studied.
The concept of static and spherically symmetric black hole solutions has been studied in $f(Q)$ gravity \cite{Black}. Also, the possibility of avoiding or at least minimizing energy conditions in the context of teleparallel gravity, has therefore brought new life to research in wormhole physics. In this view, Lin and Zhai have explored the application of $f(Q)$ gravity to the spherically symmetric configurations \cite{Zha}. They considered a simple model, $f(Q)=Q+\alpha Q^{2}$, and a polytropic EoS to investigate the internal spherically symmetric configuration. Wang et al. have studied static and spherically symmetric solutions with an anisotropic fluid for general $f(Q)$ gravity \cite{Wang}. Moreover,  Sharma et al. have investigated  the wormhole solutions in the backdrop of symmetric teleparallel gravity \cite{Sha}. By considering a known shape and redshift function, they have shown that some of the $f(Q)$ models can provide solutions which satisfy energy conditions.
In \cite{Mus}, the authors obtained wormhole solutions from the Karmarkar condition in $f(Q)$ gravity formalism. They have indicated that the combination of such ingredients provides the possibility of obtaining traversable wormholes satisfying the energy conditions \cite{Mus}. Hassan et al. have studied the traversable wormhole
geometries in $f(Q)$ by considering two specific EoS \cite{Hassan}. They have presented solutions  for a specific
shape function in the fundamental interaction of gravity (i.e., for linear form of
$f(Q)$). The presented solutions in \cite{Hassan} violate the energy conditions. In \cite{sym}, the authors discussed the existence of wormhole solutions with the help of the Gaussian and Lorentzian distributions of linear and exponential models. Along this way, Banerjee et al. have used a constant  redshift function with different shape functions to construct wormhole solutions with some known $f(Q)$ functions \cite{Ban}. They have found solutions which have not respected the energy conditions.

It should be noted that the role of  EoS in cosmology and gravity research is very important because
it describes the physical fluid which is essential to sustain a geometry. Fluid with a linear EoS (i.e., $p=\omega \rho$) and positive energy density is a good candidate to explain the evolution of the cosmos. Phantom wormholes are the famous wormhole solutions in the background of GR \cite{phantom}. Modified field equations in the context of modified theories always are more complex than GR. In this context, $f(Q)$ gravity  features in second order field equations instead of the fourth order ones in $f(R) $. Due to the complexity of the equations, finding wormhole with a liner EoS in the context of $f(R)$, $f(R,T)$, or $f(Q)$ gravity is very difficult.

 In the present paper, we study the wormhole in the background of $f(Q)$ gravity.
We focus on $f(Q)$ gravity, introduced by Jimenez et al.,\cite{Jim} where the gravitational interaction is described
by the non-metricity term $Q$.  In this view, we will investigate three types of  wormhole geometries by considering: (1)	a linear EoS between the radial pressure and energy density, (2)
isotropic wormhole i.e., $p=p_t$ (3) an asymptotically linear EoS.
  Some properties of possible solutions will be explained. The essential mathematical consideration, to construct wormhole solutions, is used to find solutions with a linear EoS. The ability of $f(Q)$ gravity, to respect energy conditions in static wormholes, is discussed. The organization of the paper is as follows:  First, we discussed conditions and equations governing wormhole then a brief review on  $f(Q)$ theory and the classical energy conditions is presented. In Sec. \ref{sec3} by defining a linear EoS, we have found some new solutions in $f(Q)$ gravity then the possibility of the isotropic solutions is investigated. Also the solutions with variable EoS are discussed in Sec.(\ref{sec3}). Solutions with variable redshift function are investigated in Sec.(\ref{sec4})  We have  shown that  some of the solutions in the $f(Q)$ gravity satisfy energy conditions at least in some regions of the space.  Finally, we have presented our concluding remarks in the last section. In this paper, We have assumed gravitational units, i.e., $c = 8 \pi G = 1$.

\section{Basic formulation of wormhole }
The line element of the general spherically symmetric wormhole is considered as:
\begin{equation}\label{1}
ds^2=-U(r)dt^2+\frac{dr^2}{1-\frac{b(r)}{r}}+r^2(d\theta^2+\sin^2\theta
d\phi^2)
\end{equation}
 where $U(r)=\exp (2\phi(r)) $. The function $ \phi(r)$ is called the redshift function which can be used to detect the redshift of the signal  by a distant observer. On the other hand, $b(r)$  is called the shape or form function. The throat of the wormhole is shown by $r_0$ which connects two universes or distinct parts of the same universe. The shape function obeys
\begin{equation}\label{2}
b(r_0)=r_0.
\end{equation}
This function describes the geometrical structure of the wormhole. Two other conditions:
\begin{equation}\label{3}
b'(r_0)<1
\end{equation}
and
\begin{equation}\label{4}
b(r)<r,\ \ {\rm for} \ \ r>r_0,
\end{equation}
are essential to construct traversable  wormholes, the former is famous as flaring out condition. We also imposed asymptotically flat condition as follow
\begin{equation}\label{5}
\lim_{r\rightarrow \infty}\frac{b(r)}{r}=0,\qquad   \lim_{r\rightarrow \infty}U(r)=1
\end{equation}
Wormhole with a constant redshift function guarantees the absence of horizon around the throat and presents zero tidal force.

Now, we present a brief review of $f(Q)$ theory of gravity. The action for symmetric teleparallel
gravity is given by
\begin{equation}\label{6}
S=\int\frac{1}{2}f(Q)\sqrt{-g}\;d^{4}x+\int L_{m}\sqrt{-g}\;d^{4}x.
\end{equation}
where $f(Q)$ is a function form of $Q$, $g$ is the determinant
of the metric, and $L_m$ is the matter Lagrangian density. The non-metricity
tensor and its trace can be defied by
\begin{equation}\label{7}
Q_{\lambda\mu\nu}=\nabla_{\lambda}g_{\mu\nu},
\end{equation}
\begin{equation}\label{8}
Q_{\alpha}=Q_{\alpha\quad\mu}^{\;\:\mu}\;\;,\;\;\widetilde{Q}_{\alpha}=Q^{\mu}_{\,\,\alpha\mu}.
\end{equation}
By using the non-metricity tensor, one can write the non-metricity conjugate
as
\begin{equation}\label{9}
P^{\alpha}_{\mu\nu}=\frac{1}{4}\left[-Q^{\alpha}_{\mu\nu}+2Q_{(\mu\quad\nu)}^{\quad\alpha}+Q^{\alpha}g_{\mu\nu}-\widetilde{Q}^{\alpha}g_{\mu\nu}-\delta^{\alpha}_{(\mu}Q_{\nu)}\right]
\end{equation}

so
\begin{equation}\label{10}
Q=-Q_{\alpha\mu\nu}P^{\alpha\mu\nu}.
\end{equation}
Also the energy-momentum tensor can be written as
\begin{equation}\label{11}
T_{\mu\nu}=-\frac{2}{\sqrt{-g}}\frac{\delta(\sqrt{-g}L_{m})}{\delta\;g^{\mu\nu}}.
\end{equation}

By varying the action (\ref{6}) with respect to metric tensor $g_{\mu\nu}$ the   motion equations
\begin{eqnarray}\label{12}
T_{\mu\nu}=-\frac{2}{\sqrt{-g}}\nabla_{\gamma}(\sqrt{-g}f_{Q}P^{\gamma}_{\mu\nu})-\frac{1}{2}g_{\mu\nu}f \nonumber \\
-f_{Q}(P_{\mu\gamma i}Q_{\nu}^{\gamma i}-2Q_{\gamma i\mu}P^{\gamma i}_{\nu}).
\end{eqnarray}
and
\begin{equation}\label{13}
\nabla_{\mu}\nabla_{\nu}(\sqrt{-g}f_{Q}P^{\gamma}_{\mu\nu})=0.
\end{equation}
are achieved where $f_Q\equiv \frac{df}{dQ}$. We consider a diagonal energy momentum tensor in the form  $T^{\mu}_{\nu}=diag[-\rho, p,p_t,p_t]$, where $\rho$ is the energy density, $p$ the radial pressure and $p_t$ the tangential pressure, respectively. Using Eq.(\ref{7}) and the line element (\ref{1}), one can show
\begin{equation}\label{14}
Q=-\frac{2}{r}\biggl(1-\frac{b(r)}{r}\biggr)\biggl(2\phi^{\prime}+\frac{1}{r}\biggr).
\end{equation}
Now, by substituting (\ref{1}) and (\ref{14}) in (\ref{12}), one can find the following field equations
\begin{eqnarray}\label{15}
\rho=&\biggl[&\frac{1}{r}\left(\frac{1}{r}-\frac{rb^{\prime}(r)+b(r)}{r^{2}}+2\phi^{\prime}(r)\left(1-\frac{b(r)}{r}\right)\right)\biggr]f_{Q} \nonumber \\
&+&\frac{2}{r}\left(1-\frac{b(r)}{r}\right)f'_{Q}+\frac{f}{2}.
\end{eqnarray}

\begin{equation}\label{16}
p=-\biggl[\frac{2}{r}\left(1-\frac{b(r)}{r}\right)\left(2\phi^{\prime}(r)+\frac{1}{r}\right)-\frac{1}{r^{2}}\biggr]f_{Q}-\frac{f}{2}.
\end{equation}

\begin{eqnarray}\label{17}
p_t(r)=&-&\biggl[(1-\frac{b(r)}{r})\left(\frac{1}{r}+\phi^{\prime}(r)(3+r\phi^{\prime}(r)) +r\phi^{\prime\prime}\right) \nonumber \\
&-&\frac{rb^{\prime}(r)
 -b(r)}{2r^{2}}(1+r\phi^{\prime})\biggr]\frac{f_{Q}}{r}\nonumber \\
&-&\frac{1}{r}(1-\frac{b(r)}{r})(1+r\phi^{\prime}(r))f'_{Q}-\frac{f}{2}.
\end{eqnarray}
Note that the prime denotes the derivative $\frac{d}{dr}$.  We have three equations, (\ref{15})-(\ref{17}),
six unknown functions, namely, $f(Q)$, $b(r)$, $\phi(r)$, $\rho(r)$, $p(r)$ and $p_t(r)$. In Sec.(\ref{sec3}), for the sake of simplicity, we set $\phi(r)=0$ so to close the system one may add  an extra EoS or choose some of the unknown functions arbitrarily. Vanishing redshift functions guaranties the absence  of horizons and leads to vanishing tidal force.
In the recent part of this paper, we will discuss some mathematical methods which have been used in the literature to find exact wormhole solutions then try to use some of theses algorithms and  present some new wormhole solutions in the context of $f(Q)$ gravity.

Let us devote some words to classical energy conditions, the  null energy condition (NEC), dominant energy condition (DEC),weak energy condition (WEC), and strong energy condition (SEC), are defined as:
\begin{eqnarray}\label{18}
\textbf{NEC}&:& \rho+p\geq 0,\; \rho+p_t\geq 0 \\
\label{19}
\textbf{WEC}&:& \rho\geq 0,\; \rho+p\geq 0,\;\rho+p_t\geq 0, \\
\textbf{DEC}&:& \rho\geq 0, \;\rho-|p|\geq 0,\;\rho-|p_t|\geq 0, \\
\textbf{SEC}&:& \rho+p\geq 0,\, \;\rho+p_t\geq 0,\;\rho+p+2p_t \geq 0. \label{20}
\end{eqnarray}
Wormhole violates these energy conditions in the context of GR. We will analyze these conditions by defining
\begin{eqnarray}\label{21}
 H(r)&=& \rho+p ,\, H_1(r)= \rho+p_t,\, H_2(r)= \rho-|p|, \nonumber \\
 H_3(r)&=&\rho-|p_t|,\, H_4(r)= \rho+p+2p_t,
\end{eqnarray}
 for our solutions.

 In \cite{vol}, a volume integral
\begin{eqnarray}\label{22}
I_V\equiv \int (\rho+p_r)dV&=&\left[ (r-b)\ln \left(\frac{e^{2\phi(r)}}{1-b/r}\right)\right]^\infty_{r_0}
   \nonumber  \\
&& \hspace{-1.7cm} - \int_{r_0}^\infty
(1-b'(r))\left[ \ln \left(
\frac{e^{2\phi(r)}}{1-\frac{b}{r}}\right)\right] dr.
\end{eqnarray}
has been introduced by Visser et. al.   to measure the amount of NEC violation. However, Zaslaveskii \cite{zas} has introduced another  integral in which  the wormhole spacetime can be divided into two different regions; the region filled by exotic matter i.e., $r_0<r\leq a$, and the region filled by ordinary matter  i.e., $r\geq a$, \cite{variable}. So, the  total amount of exotic matter is measured by
\begin{equation}\label{23}
I=8\pi \int_{r_0}^a (\rho+p)r^{2}dr .
\end{equation}

\section{Wormhole solutions with constant redshift function }\label{sec3}
 In this section, we study some new asymptotically flat wormhole solutions and discuss some of their physical properties. We will investigate the mathematical considerations to construct exact analytical wormhole solutions in $f(Q)$ gravity. We analyze the energy conditions for all of the solutions. In the recent part of this paper, we set $r_0 =1$.

\subsection{Wormholes with  linear equation of state}\label{subsec1}
Considering an EoS, which present a relation between the energy momentum tensor components  then solving the field equations to construct wormhole solutions, is the most acceptable technique to find wormhole solutions. This method has been used extensively in GR to find exact wormhole solutions. In the modified theories of gravity, the complexity of field equations usually does not allow the researchers to find analytical exact solutions so the numerical methods are used or the EoS is not considered. In this case, $f(Q)$ gravity also features in second order field equations instead of the fourth order ones in $f(R)$ \cite{Zha}.
 Cosmos with a linear EoS is the most considerable one which have been investigated in the literature. A linear relation, $p=\omega \rho$, is the most usual EoS which has been used in the literature for studying the Cosmos or finding wormhole solutions. In this section, we will investigate some wormhole solutions with an EoS in the form
\begin{equation}\label{24}
p(r)= \omega \rho(r)
\end{equation}
where $\omega$ is the EoS parameter. Asymptotically flat  wormhole solutions with $\omega\leq -1$, dubbed phantom EoS, have been studied in \cite{phantom1}. As we know, there is not any presented asymptotically flat wormhole solution, with linear Eos, in the literature.  In \cite{Hassan}, it was mentioned that finding wormhole solutions with linear EoS is very difficult but we will try to find some exact wormhole solutions. To complete the algorithm, one should consider a $f(Q)$ function then by using Eq.(\ref{15}) and (\ref{24}) close the system. Considering Eqs.(\ref{15}) and (\ref{24})
 yields
 \begin{eqnarray}\label{25}
&\omega&\biggl( \left [\frac{1}{r^2}-\frac{rb^{\prime}(r)+b(r)}{r^{3}}\right ]f_{Q}+\frac{2}{r}\left(1-\frac{b(r)}{r}\right)f'_{Q}+\frac{f}{2}\biggr)= \nonumber \\
&-&\left[\frac{2}{r^2}\left(1-\frac{b(r)}{r}\right)-\frac{1}{r^{2}}\right]f_{Q}-\frac{f}{2}.
\end{eqnarray}
Solving this equation for a general function of $f(Q)$ is very difficult so we will use some special functions for $f(Q)$ to find $b(r)$. In particular, we have used the following functions
\begin{equation}\label{26}
f(Q)=exp(Q)
\end{equation}
\begin{equation}\label{27}
f(Q)=-ln(1-Q)
\end{equation}
\begin{equation}\label{28}
f(Q)=\alpha (-Q)^{n}+C
\end{equation}
\begin{equation}\label{29}
f(Q)=\frac{1}{1-Q}
\end{equation}
but finding  the exact wormhole solutions is formidable for all of these functions.
Finally, we consider
\begin{equation}\label{30}
f(Q)=\alpha (-Q)^{n}
\end{equation}
where $\alpha$ and $n$ are some arbitrary constant. It should be noted that since $b(r)<r$, the $Q(r)$ function is always negative therefore  the minus sign appears in (\ref{30}).
By using the functions (\ref{14}), (\ref{30}), and Eq.(\ref{25}), one can find a solution for the shape function as follow
\begin{equation}\label{31}
b(r):=C_0  r^{C_1(n,\omega)}+C_2(n,\omega) r
\end{equation}
where
 \begin{equation}\label{32}
C_1(n,\omega)=\frac{6n^2 \omega-7n\omega+2n+\omega+1}{n\omega(2n-1)}
\end{equation}
 \begin{equation}\label{33}
C_2(n,\omega)=\frac{(n-1)(4n\omega-\omega-1)}{4n^2 \omega-6n \omega-2n+\omega+1}
\end{equation}
and $C_0$ is the constant of integration. In order to have an asymptotically flat solution, one can impose
 \begin{equation}\label{34}
C_2(n,\omega)=0
\end{equation}
 which yields
  \begin{equation}\label{35}
n=\frac{\omega+1}{4\omega}
\end{equation}
or
 \begin{equation}\label{36}
\omega=\frac{1}{4n-1}
\end{equation}
also shape function with $n=1$ can satisfy (\ref{25}) but this  leads to$f(Q)=Q$ which recovers the GR.
Using (\ref{32}) and (\ref{36}) gives
\begin{equation}\label{37}
b(r)=r^{\gamma}
\end{equation}
where
\begin{equation}\label{38}
\gamma=-\frac{2n+1}{2n-1}.
\end{equation}
Due to asymptotically flat condition, $\gamma<1$ is acceptable which shows that $n$ in the range
\begin{equation}\label{39}
0\leq n \leq1/2
\end{equation}
is not allowed. Equations (\ref{39}) and (\ref{36}) demonstrate that
\begin{equation}\label{40}
-1< \omega <1
\end{equation}
is the possible range for EoS parameter.
Now, we can see that for any acceptable $\omega$ and $n$ one can reach a shape function in the power-law form, $b(r)=r^ \gamma$. This result can be explained in a reverse way; The power-law shape function and $f(Q)$ function can lead to a wormhole solution with a special EoS parameter. The power-law shape function is the most famous one in all classes of wormhole theories. To proceed further, we have assumed some special cases for solutions. The case, $n=-1$ yields, $\omega=-1/5$ which implies
\begin{equation}\label{41}
f(Q)=\frac{\alpha}{-Q}
\end{equation}
Using the field equation (\ref{15}) shows that energy density diverges, as $r\rightarrow \infty $, so this solution is not  physically interesting. As the second case, we set $n=3/2$ which leads to $\omega=1/5$,  $b(r)=1/r^2$ and $f(Q)=\alpha (-Q)^{3/2}$. By using field equations, one can find
\begin{equation}\label{42}
p(r)=\frac{1}{5}\rho(r)= \frac{1}{5}\sqrt{\frac{25(r^3-1)(r^3-4)^2}{2r^{10}}}.
\end{equation}

\begin{figure}
\centering
  \includegraphics[width=3 in]{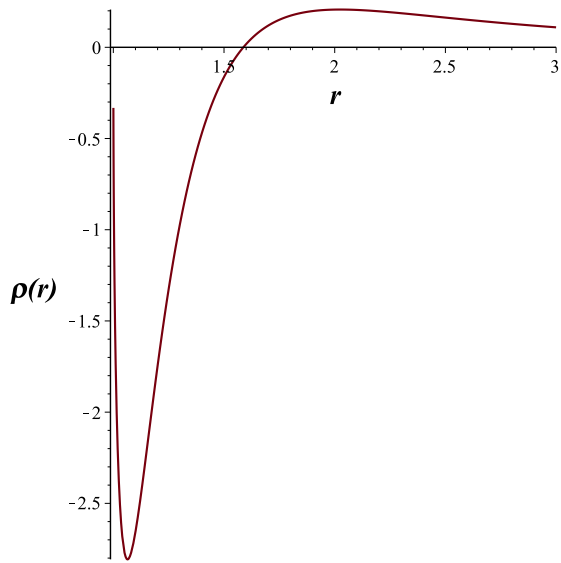}
\caption{The figure represents the $\rho(r)/\alpha$ against radial coordinate for $\alpha=1$, $n=3/2$, and $\gamma=-2$  which is positive for $r<r_1$ and negative for $r_1<r$. See the text for details.}
 \label{fig1}
\end{figure}
\begin{figure}
\centering
  \includegraphics[width=3 in]{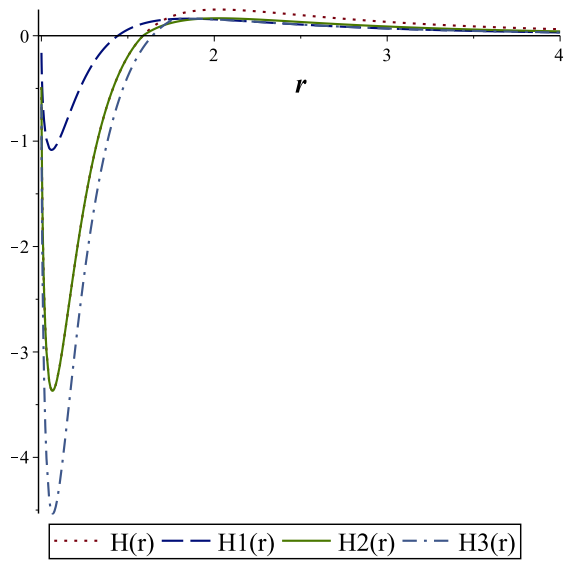}
\caption{The graphical behavior of $H(r)$ (dotted line),$H_1 (r)$ (dashed line),$H_2 (r)$ (solid line), and $H_3 (r)$ (dashed-dotted line) against $r$ for the case $\alpha=-1$, $n=3/2$, and $\gamma=-2$.  It is clear that NEC, WEC, and DEC is violated only near the throat. See the text for details.}
 \label{fig2}
\end{figure}

\begin{figure}
\centering
  \includegraphics[width= 3 in]{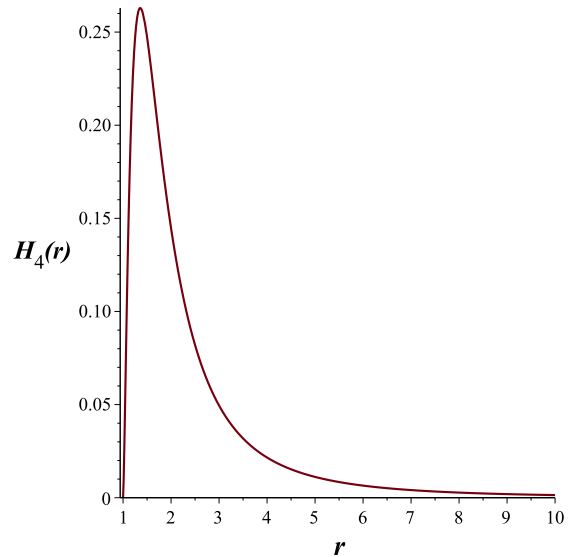}
\caption{The plot depicts the general behavior of $H_4$(r) against $r$ for the case $\alpha=-1$, $n=3/2$, and $\gamma=-2$. It is clear that $H_4(r)$ is  positive in the entire range of $r$. It shows that the third condition of SEC is satisfied.
 See the text for details.}
 \label{fig3}
\end{figure}

\begin{figure}
\centering
  \includegraphics[width=3 in]{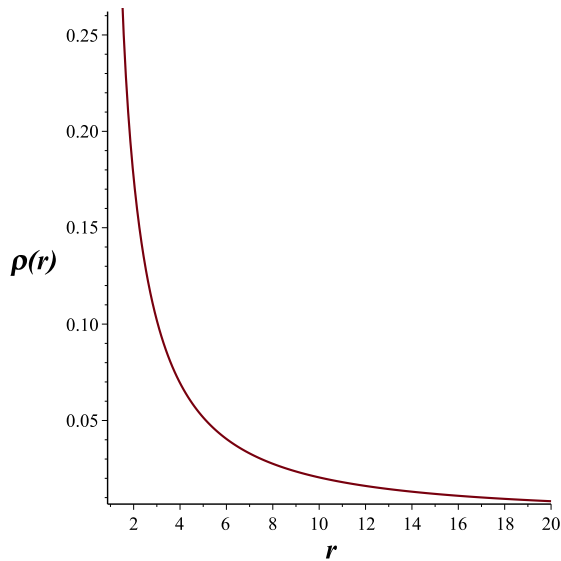}
\caption{The figure represents the $\rho(r)$ against radial coordinate for $\alpha=1$, $n=2/3$, and $\gamma=-7$  which is positive for the entire range. See the text for details.}
 \label{fig4}
\end{figure}

The $\rho(r)/\alpha$ is depicted as a function of radial coordinate in Fig. \ref{fig1}. This figure indicates that $\rho$ is negative in the vicinity of throat but for $r>r_1$ is positive while $\alpha> 0$ is considered. It is clear that for $\alpha< 0$, the energy density is positive in the vicinity of the throat and for $r>r_1$ is negative. For $\alpha=-1$, we have plotted $H(r), H_1(r), H_2(r)$, and $H_3(r)$ as a function of $r$ in Fig. \ref{fig2}.  It can be deduced from Fig. \ref{fig2} that the violation of NEC, WEC and DEC is restricted to some regions near the throat. We have plotted $H_4 (r)$ as a function of $r$ in Fig. \ref{fig3} which shows that third condition for SEC is  satisfied  the whole spacetime so one can conclude that SEC is violated only in the vicinity of the throat.

\begin{figure}
\centering
    \includegraphics[width=3 in]{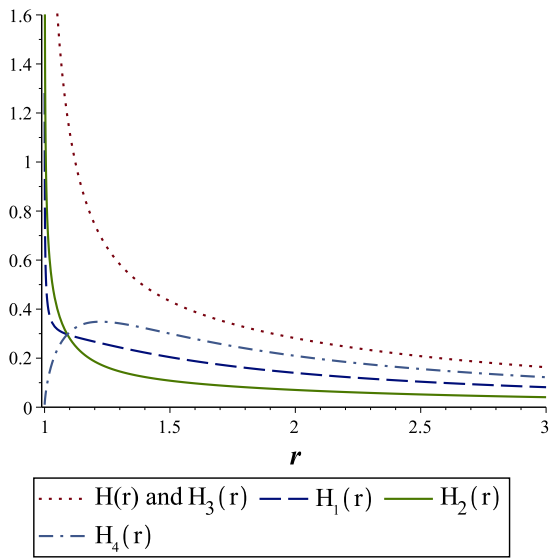}
\caption{The graphical behavior of $H(r)$ (dotted line),$H_1 (r)$ (dashed line),$H_2 (r)$ (solid line),  $H_3 (r)$ (dotted line), and $H_4 (r)$ (dashed-dotted line),  against $r$ for the case $\alpha=1$, $n=2/3$, and $\gamma=-7$. Since all of these functions are positive,   all the energy conditions are respected in the whole  space. See the text for details.}
 \label{fig5}
\end{figure}

\begin{figure}
\centering
  \includegraphics[width=3 in]{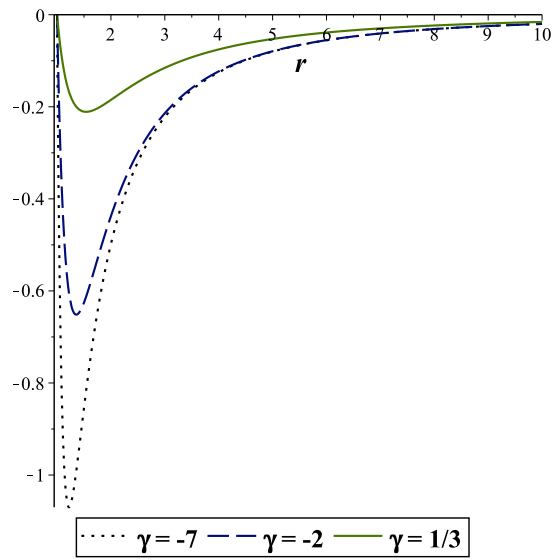}
\caption{The plot depicts the  behavior of $Q(r)$ against $r$ for $\gamma=-2$ (dotted line), $\gamma=-7$ (dashed line), and $\gamma=1/3$ (solid line). One can see that $Q(r)$ is a vanishing function at the boundaries. See the text for details.}
 \label{fig6}
\end{figure}

In the last choice, we set $n=2/3$ which leads to $\omega=3/5$, $b(r)=1/r^7$, and $f(Q)=\alpha (-Q)^{2/3}$. It is easy to show that
 \begin{equation}\label{43}
p(r)=\frac{3}{5} \rho(r)=\frac{3}{8}\frac{2^{2/3}(r^8+1)}{r^{10}(\frac{r^8-1}{r^{10}})^{1/3}},
\end{equation}
for $\alpha=-1$. We have plotted $\rho(r)$ as a function of $r$ in Fig. \ref{fig4}. It can be concluded that energy density is positive in the whole space. Also, we have plotted $H(r), H_1(r), H_2(r)$, $H_3(r)$ and $H_4(r)$ as a function of $r$ in Fig. \ref{fig5}. Figures \ref{fig4} and \ref{fig5} verified that all the energy conditions are respected in the whole  space for this solution. Now, let us discuss  the  behavior of the energy momentum tensor at boundaries. It is easy to show that
\begin{equation}\label{44}
\lim_{r\rightarrow \infty}\rho(r)=0.
\end{equation}
\begin{equation}\label{45}
\lim_{r\rightarrow 1}\rho(r)\rightarrow \infty.
\end{equation}
Let us devote some words to divergence of energy density on the boundary conditions,  we have plotted $Q(r)$ as a function of $r$ in Fig. \ref{fig6} for three shape function in the form of Eq.(\ref{37}) with $\gamma=-2$,$\gamma=-7$ and $\gamma=1/3$ respectively. One can deduce from Fig. \ref{fig6}  that $Q(r)$ is vanishing at the throat and large infinity distance. Due to the functional form of $f(Q)$ (see Eq.(\ref{30})), it can be concluded that this function or its first derivative  will diverge for $n<1$ at $Q=0$. Because of the aforementioned reason, the energy density tensor components may diverge for some ranges of $n$.

\subsection{Isotropic wormholes }\label{subsec2}

Usually the energy momentum tensor of the asymptotically flat  wormholes in the GR are anisotropic i.e. $p(r)\neq p_t(r)$.  In this section, we consider an EoS as follow,
 \begin{equation}\label{46}
p(r)=p_t (r).
\end{equation}
Then try to find wormhole solutions with some different form of $f(Q)$. For the case $f(Q)=Q^n$, Eq.(\ref{46}) leads to
\begin{equation}\label{47}
b(r)=\frac{2(n-1)}{2n-1}r+C r^3,
\end{equation}
where $C$ is integration constant. It is clear that this solution is not asymptotically flat. Using the same algorithm for $f(Q)$ in the form of (\ref{26}) gives
\begin{equation}\label{48}
b(r)=-\frac{1}{4} r^3+r\pm \frac{\sqrt{8Cr^4+r^4-8r^2}}{4}
\end{equation}
which is not asymptotically flat. Also, we have found that the solutions for $f(Q)$ models (\ref{27}) and (\ref{28})  are not asymptotically flat.
It seems that finding asymptotically flat wormhole solutions, with constant redshift function and isotropic pressure, in $f(Q)$ scenario is formidable or at least very difficult.
\begin{figure}
\centering
  \includegraphics[width=3 in]{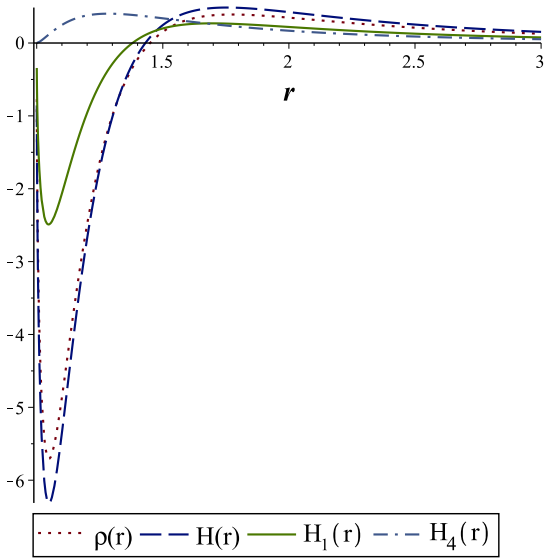}
\caption{The graphical behavior of $\rho(r)$ (dotted line), $H (r)$ (dashed line), $H_1 (r)$ (solid line),  and $H_4 (r)$ (dashed-dotted line),  against $r$ for the case $\alpha=1$, $n=3/2=$, and $\gamma=-4$.
It shows that energy conditions are violated only near the wormhole throat. See the text for details.}
 \label{fig7}
\end{figure}

\subsection{Asymptotically linear equation of state }\label{subsec3}
As it was mentioned, finding wormhole solutions for our presented models of $f(Q)$ is formidable except for (\ref{30}) under some mathematical considerations. In this section, we use the algorithm which has been used in \cite{variable} to construct wormhole with asymptotically linear EoS.  One may postulate that  ,the EoS of fluid supporting wormhole near the throat, has a  different behavior from those of the Universe in the large scale but local behavior of fluid near the throat will tend
to a global EoS in the large scale \cite{variable}. This seems to be a good motivation to construct wormhole with variable EoS parameter which tends to a constant at large scales. In this case, the EoS parameter is given by
\begin{equation}\label{49}
\omega(r)=\omega_\infty +g(r).
\end{equation}
where $\omega_\infty$ is the EoS at large distance from the throat and $g(r)$ should admits
\begin{equation}\label{50}
\lim_{r\rightarrow \infty}g(r)=0.
\end{equation}
\begin{figure}
\centering
  \includegraphics[width=3 in]{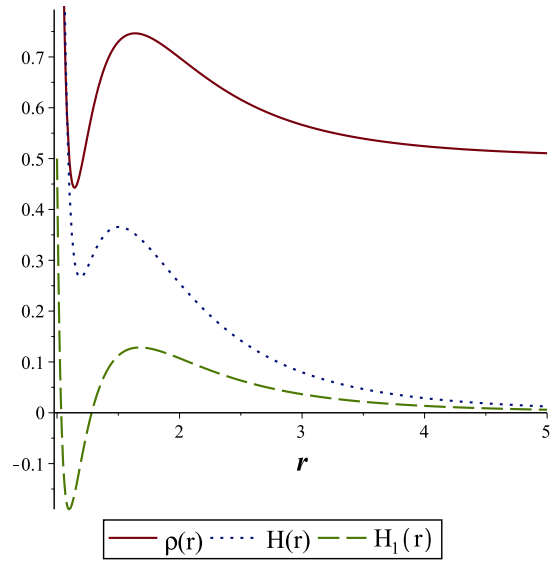}
\caption{The graphical behavior of $\rho(r)$ (solid line), $H (r)$ (dotted line), $H_1 (r)$ (dashed line) against $r$ for the case $f(Q)=\exp (Q)$, and $\gamma=-2$. It shows that the second  NEC is violated only in a small region of the space while it is respected at the throat and large infinity distance . See the text for details.}
 \label{fig8}
\end{figure}
\begin{figure}
\centering
  \includegraphics[width=3 in]{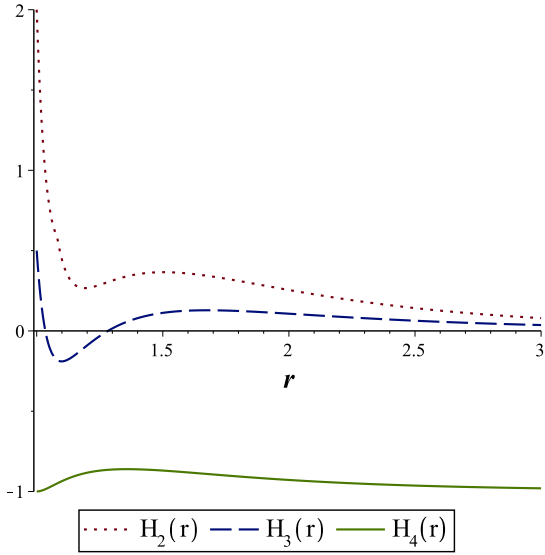}
\caption{The figure represents the $H_2(r)$ (dotted line), $H_3(r)$ (dashed line) and $H_4(r)$ (solid line) against radial coordinate for $f(Q)=\exp (Q)$, and $\gamma=-2$ which shows $H_2 (r)$ is always positive. It is clear that  $H_4 (r)$ is negative in the whole space while $H_3(r)$ is negative in a small region.
  positive for the entire range. See the text for details.}
 \label{fig9}
\end{figure}

 \begin{figure}
\centering
  \includegraphics[width=3 in]{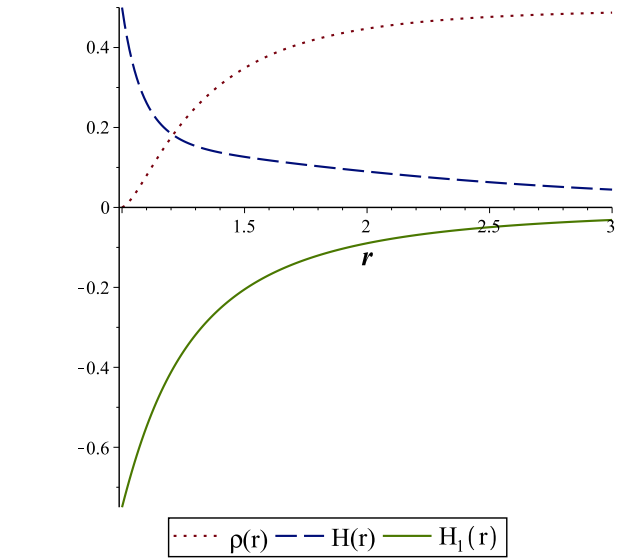}
\caption{The plot depicts  $\rho(r)$ (dotted line), $H(r)$ (dashed line) and $H_1(r)$ (solid line) against radial coordinate for $f(Q)=1/(1-Q)$, and $\gamma=1/2$ which shows $\rho(r)$ and $H(r)$ are always positive but $H_1(r)$ is always negative. One can concluded that WEC and NEC are violated everywhere.  See the text for details.}
 \label{fig10}
\end{figure}
\begin{figure}
\centering
  \includegraphics[width=3 in]{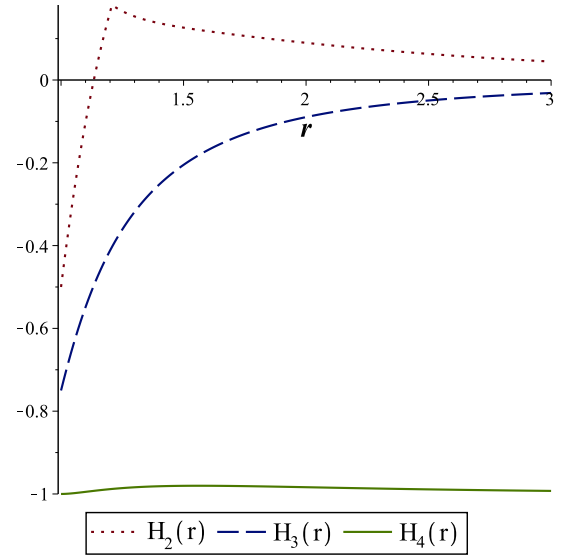}
\caption{The graphical behavior of $H_2(r)$ (dotted line), $H_3 (r)$ (dashed line), $H_4 (r)$ (solid line) against $r$ for the case $f(Q)=1/(1-Q)$, and $\gamma=1/2$. It shows that DEC and SEC are violated everywhere. See the text for details.}
 \label{fig11}
\end{figure}

\begin{figure}
\centering
  \includegraphics[width=3 in]{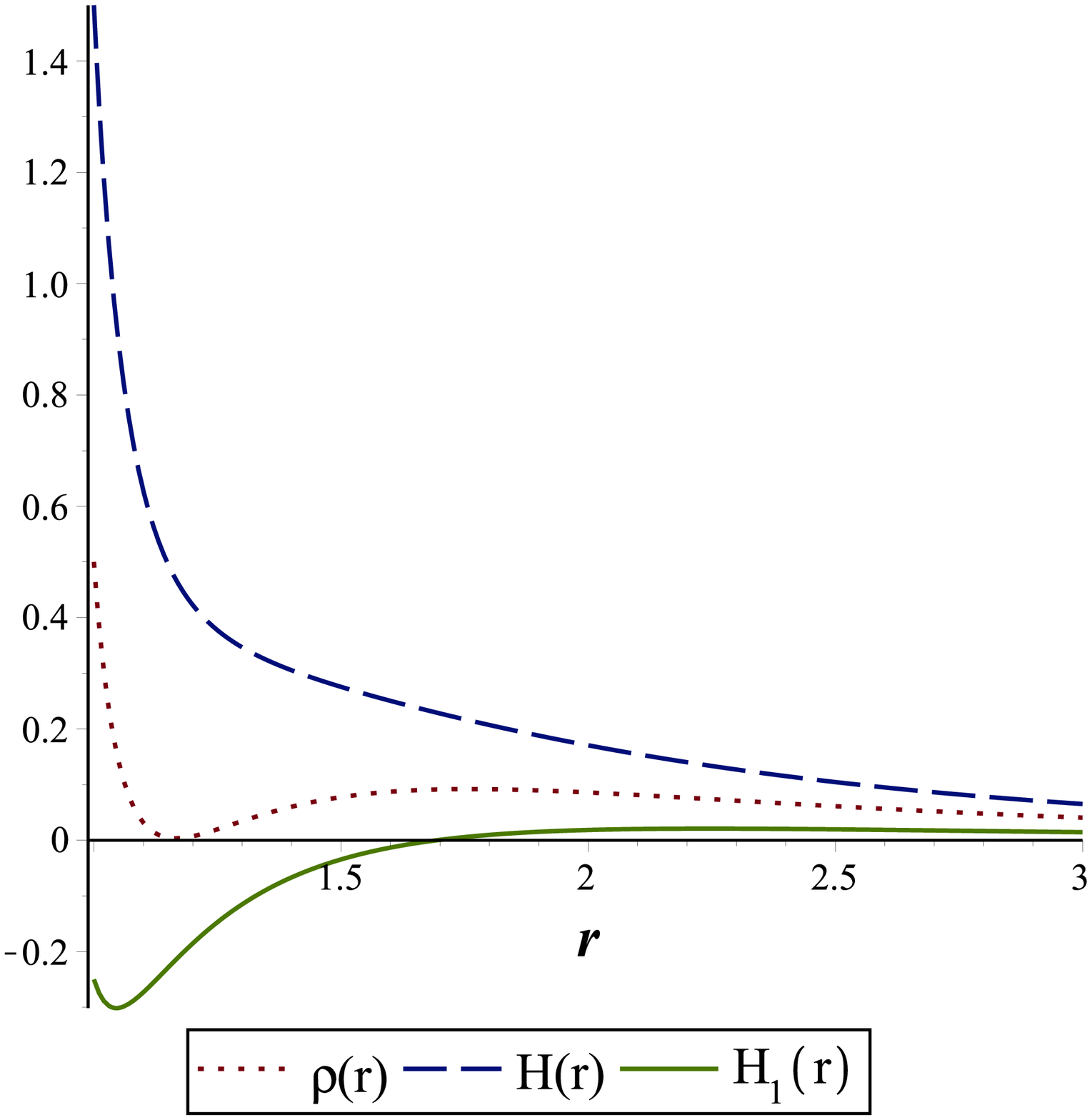}
\caption{The plot depicts  $\rho(r)$ (dotted line), $H(r)$ (dashed line) and $H_1(r)$ (solid line) against radial coordinate for $f(Q)=-\ln(1-Q)$, and $\gamma=-1/2$ which shows $\rho(r)$ and $H(r)$ are always positive but $H_1(r)$ is  negative in the vicinity of throat. One can concluded that WEC and NEC are respected everywhere except a small region in the vicinity of throat.  See the text for details.}
 \label{fig12}
\end{figure}
So the EoS i.e., $p=\omega(r) \rho$, is asymptotically linear.
Now, we can investigate solutions with variable EoS in the context of $f(Q)$ gravity. As the first example, we consider $f(Q)$ in the form of Eq.(\ref{30}) and the shape function (\ref{37}) then by using Eqs.(\ref{15}) and (\ref{16}), after some straightforward calculation, one can show
\begin{equation}\label{51}
\omega(r)=\frac{1}{4n-1}+g(r)
\end{equation}
which $g(r)$ is a function of $r$ that obeys condition (\ref{50}). In particular, $\gamma=-4$ gives
\begin{equation}\label{52}
g(r)=\frac{1-2n}{r^5(4n^2-5n+1)-14n^2+11n+1}
\end{equation}
and obeys condition (\ref{50}). A large number of solutions can be presented for different values of $n$ and $\gamma$. One should note that for the case, $\gamma=-\frac{2n+1}{2n-1}$, the solution reduced to a linear EoS. We have plotted the $\rho, H(r), H_1(r)$ as a function of $r$ in Fig. \ref{fig7} which generally present the same behavior in contrast to linear EoS solutions in Fig. \ref{fig2}.
It was mentioned that finding wormhole solutions with linear EoS is very difficult or formidable for the models (\ref{26}- \ref{29}) so we try to find some asymptotically linear EoS solutions for these models. If we consider (\ref{26}) and (\ref{37}), it is easy to show that
\begin{equation}\label{53}
\omega(r)=-1+g_1(r)
\end{equation}
which
\begin{eqnarray}\label{54}
g_1(r)&=&(-2+4r^{m-1})  [ r^6-(2m+2)r^{m+3}+2r^4  \nonumber \\
&-&8r^{2m}(m-3)+8r^{m+1}(m-5)+16r^2  ] ^{-1}.
\end{eqnarray}

This can be considered as an exact wormhole solution with variable EoS for $f(Q)=\exp (-Q)$. We have plotted $\rho, H(r)$ and $H_1(r)$ as a function of $r$ in Fig. \ref{fig8} which indicates that energy density and $H(r)$ is positive in the entire range of spacetime but the second NEC i.e., $H_1(r)\geq 0$, is violated only in a small region of the spacetime. Further, we have plotted $H_3(r)$ and $H_4 (r)$ in Fig. \ref{fig9}. One can deduce from Fig. \ref{fig9} that DEC is respected at the throat and large distance in the other word DEC is violated  only in a small region $r_1<r<r_2$. Also, it is clear from Fig. \ref{fig9} that SEC is violated in the whole spacetime. Besides, one can show that
\begin{equation}\label{55}
\lim_{r\rightarrow \infty}\frac{p(r)}{p_t (r)}=1
\end{equation}
 which implies that this solution is asymptotically isotropic. It seems that this solution provides many physical conditions to be accepted as a possible solution in $f(Q)$ gravity.  We can show that for $f(Q)=\frac{1}{1-Q}$ and power-law shape function i.e. $b(r)=1/r^2$, EoS parameter is

\begin{figure}
\centering
  \includegraphics[width=3 in]{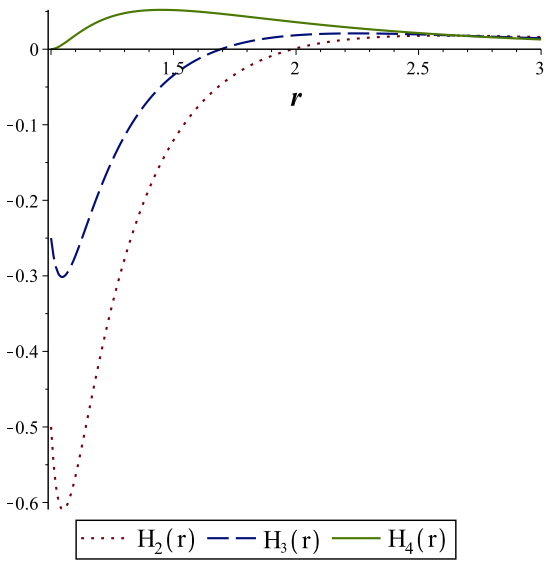}
\caption{The plot depicts  $H_2(r)$ (dotted line), $H_3(r)$ (dashed line) and $H_4(r)$ (solid line) against radial coordinate for $f(Q)=-\ln(1-Q)$, and $\gamma=-1/2$ which shows $H_4(r)$ is always positive but $H_2(r)$ and $H_3(r)$ are  negative in the vicinity of throat. One can concluded that DEC and SEC are respected everywhere except a small region in the vicinity of throat.  See the text for details.}
 \label{fig13}
\end{figure}
  \begin{eqnarray}\label{56}
\omega(r)&=&-1 \nonumber \\
&+&\frac{6r^3-8r^{5/2}+8r-20r^{1/2}+12}{-r^5-6r^3+7r^{5/2}-40r+90r^{1/2}-50}.
\end{eqnarray}
We have plotted $\rho, H(r)$, and $H_1(r)$ as a function of $r$ in Fig. \ref{fig10} which indicates that energy density and $H(r)$ is positive in the entire range of spacetime but the second NEC i.e., $H_1(r)\geq 0$. is not respected in the whole spacetime. Further, we have plotted $H_2(r)$, $H_3(r)$, and $H_4 (r)$ in Fig. \ref{fig11}. One can deduce from Fig. \ref{fig11} that DEC and SEC are violated in the whole spacetime. Also, it is easy to show that this solution  is asymptotically isotropic.
Finally, for the case, $f(Q)=-\ln (1-Q)$  and shape function (\ref{37}),
 \begin{equation}\label{57}
\omega(r)=-\frac{1}{\gamma}+g_2 (r)
\end{equation}
 is achieved where $g_2 (r)$ admits condition (\ref{50}).
  Since $\gamma<1$, Eq.(\ref{57}) shows that $\omega_\infty <1$ is acceptable. Let us explore the energy conditions: $\rho, H(r)$ and $H_1(r)$  are depicted in  Fig. \ref{fig12} as a function of $r$  also $H_2 (r)$, $H_3(r)$, and $H_4 (r)$ are depicted in Fig. \ref{fig13} respectively.  Figures \ref{fig12} and \ref{fig13} demonstrate that energy density, $H(r)$, and $H_4 (r)$  are positive in the entire range of spacetime while $H_1 (r)$, $H_2(r)$, and $H_3 (r)$ are negative in the vicinity of throat. Using volume integral in the form (\ref{23}) can help us to minimize the usage of exotic matter. It should be noted that in the last solutions, all of the energy conditions can be respected in  the whole space except in a small region in the vicinity of throat while in the previous solutions at least one of the energy conditions (SEC) violated in the whole spacetime.
\begin{figure}
\centering
  \includegraphics[width=3 in]{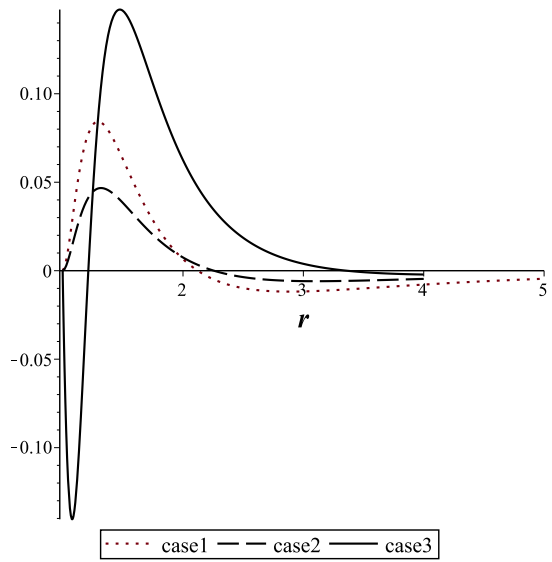}
\caption{The plot depicts the energy density for  $b(r)=1$,$f(Q)=ln(1-Q)$ and $\phi(r)=ln(\frac{\sqrt{1+r^{2}}}{r})$  (dotted line), $\phi(r)=ln(1+1/r)$ (dashed line), $\phi(r)=1/r$  (solid line) against radial coordinate  which shows $\rho(r)$ is not always positive . See the text for details.}
 \label{fig14}
\end{figure}

  \section{Solutions with variable redshift function }\label{sec4}
Till now, we have chosen $\phi'=0$ and solved the equations to find solutions. In this section, we try to investigate solutions with a non-constant redshift function. There are so many algorithms to find exact wormhole solutions while a non-constant redshift function is considered. The form of field equations in the background of $f(Q)$ gravity for variable redshift function is more complex than the vanishing redshift function. One can choose an arbitrary redshift function then try to find $b(r)$ by solving equations for different models of $f(Q)$. We have examined this algorithm for the following redshift functions,
  \begin{equation}\label{26a}
\phi(r)=ln(1+1/r)
\end{equation}
 \begin{equation}\label{27a}
\phi(r)=1/r
\end{equation}
\begin{equation}\label{28a}
\phi(r)=ln(\frac{\sqrt{1+r^{2}}}{r}).
\end{equation}

\begin{figure}
\centering
  \includegraphics[width=3 in]{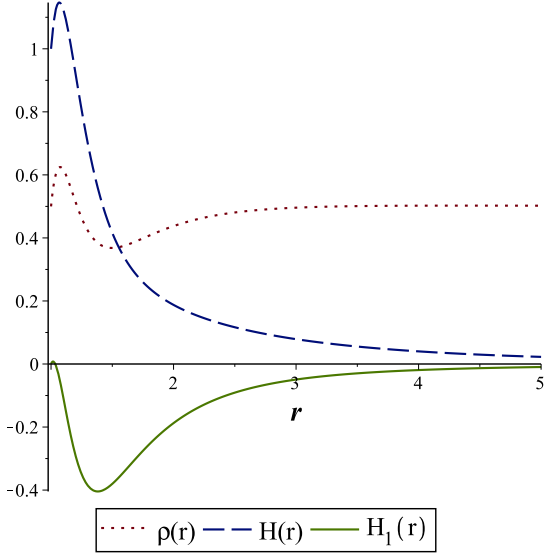}
\caption{The plot depicts $\rho(r)$ (dotted line), $H(r)$ (dashed line) and $H_1(r)$ (solid line) against radial coordinate for $b(r)=1$,$f(Q)=\exp(Q)$, and $\phi(r)=1/r$ which shows $\rho(r)$ and $H(r)$ are always positive but $H_1(r)$ is  negative. One can conclude that NEC, WEC, and SEC  are violated.  See the text for details.}
 \label{fig15}
\end{figure}
\begin{figure}
\centering
  \includegraphics[width=3 in]{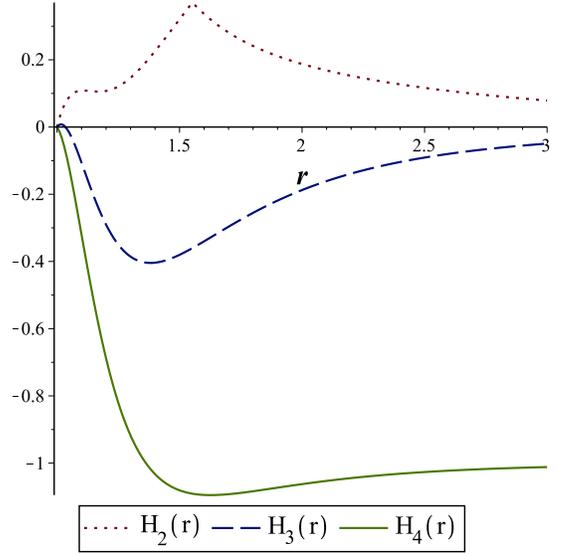}
\caption{The plot depicts  $H_2(r)$ (dotted line), $H_3(r)$ (dashed line) and $H_4(r)$ (solid line) against radial coordinate for $b(r)=1$,$f(Q)=\exp(Q)$, and $\phi(r)=1/r$ which shows $H_2(r)$ is always positive but $H_3(r)$ and $H_4(r)$ are  negative in the whole spacetime. One can conclude that DEC and SEC are violated everywhere. See the text for details.}
 \label{fig16}
\end{figure}
The result implies that analytical solutions with linear EoS  are formidable. In the next step, we used the above redshift function with some known shape functions to analyze the wormhole in the background of $f(Q)$. The energy density for the shape functions $b(r)=1, b(r)=1/\sqrt{r}$, and the redshift functions (\ref{26a}-\ref{28a}), are investigated in the background of different $f(Q)$ functions. For example, we have plotted $\rho(r)$ as a function of $r$ in Fig. \ref{fig14} for  functions, $b(r)=1$,$f(Q)=ln(1-Q)$ and the redshift functions $\phi(r)=ln(\frac{\sqrt{1+r^{2}}}{r})$ (case 1), $\phi(r)=ln(1+1/r)$ (case 2), $\phi(r)=1/r$ (case 3). This figure demonstrates that related energy densities are not a positive function everywhere; so, these solutions cannot be considered as the suitable physical solutions. As another example, for $b(r)=1$ or $b(r)=1/\sqrt{r}$ and $f(Q)=(-Q) ^{3/2}$  all of the redshift functions (\ref{26a}-\ref{28a}) lead to a energy density in the form of Fig. \ref{fig14}. On the other hand, for the case $b(r)=1/\sqrt{r}$,$f(Q)=-ln(1-Q)$, and the redshift functions (\ref{26a}-\ref{28a})  the energy density is positive everywhere but the EoS  is not asymptotically linear.
For some other cases, i.e., $b(r)=1$,$f(Q)=\exp(Q)$, and $\phi(r)=1/r$ the energy density is positive in the whole  spacetime  and the EoS  is  asymptotically linear. So, we have plotted $\rho(r), H(r)$ and $H_{1}(r)$ as a function of radial coordinate in Fig. \ref{fig15}. Also, in Fig. \ref{fig16} $H_{2}(r), H_{3}(r)$ and $H_{4}(r)$ as a function of radial coordinate are plotted. Figures \ref{fig15} and \ref{fig16} indicate that all of the energy conditions violate everywhere in spacetime. As the last example, we  have analyzed solutions with  $b(r)=1/\sqrt{r}$,$f(Q)=\exp(Q)$  and the redshift functions (\ref{26a}-\ref{28a}). We have found that these are solutions with positive energy density and asymptotically isotropic linear EoS, which respect some of the energy conditions near the wormhole throat.  These results demonstrate that finding a physical solution with a variable redshift function which has the consistency with physical conditions may be possible. Besides that, solutions with the same $f(Q)$ and shape functions provide different results with variable or constant redshift function.
 To summarize, a set of variable redshift functions, which are famous in the literature, are investigated in the context of $f(Q)$. It seems that the complexity of the equations, does not allow us for finding analytical linear exact solutions, but the numeric solutions may be accessible. Although the variable redshift function provides more solutions, but the complexity of the equations restricts our ability to find suitable analytical solutions in a general form.

\section{Concluding remarks}
Wormholes act as a short-cut way between two distant universes  or two points in a universe and can be used as a time machine.
Wormholes are the theoretical solutions of Einstein field equations, which violate energy conditions in the context of GR, so modifications of GR attracted special attention  to investigate these fantastic solutions.  These studies may open new windows to observe or construct wormhole. The main advantage of using a modified theory, in studying wormhole, is to explore the possibility of avoiding the violation of energy conditions. The exotic matter is a problematic issue so  many arguments have been presented  in favor of the violation of the energy conditions.

In this paper, we have analyzed the ability of the newly proposed theory, $f(Q)$ gravity, to describe wormhole solutions.
 The $f(Q)$ theory is used in explaining the cosmic expansion and other related concepts as well as wormhole and black hole. Since $f(Q)$ gravity is a novel approach, it can bring new insights into cosmology and gravity. We have considered four  special models for $f(Q)$ function then some asymptotically flat wormhole solutions are presented. Actually, we can divide our solutions in two classes: (i) wormholes with linear EoS, and (ii) asymptotically linear EoS. For the latter, the EoS is not locally linear in the vicinity of the throat, but is compatible  with cosmos EoS on the large scale.
   Although, modified field equations are more complex than GR, but solutions with a linear EoS, which is the most famous EoS  in the cosmology, are obtained. We have considered a variety of the models, but  finding exact wormhole solutions is formidable for linear EoS. Finally, a special form of $f(Q)=\alpha (-Q)^{n}$ is assumed and the solutions with linear EoS are obtained. We found that the shape function is a power-law function.  Using the asymptotically flat condition, a relation between $n$ and $\omega$ is found. We have shown that in some of the solutions, energy density diverges at the boundaries, which are not physically interesting. In this case, all of the energy conditions can be respected in the whole spacetime. For other solutions in this class, we  have shown that the violation of NEC, DEC, and SEC can be respected in some regions in the framework of symmetric teleparallel gravity $f(Q)$. This result is not possible in the context of GR.  Moreover, it was shown that finding an asymptotically flat wormhole with isotropic fluid is formidable or at least very difficult in $f(Q)$ gravity.

 Furthermore, we have also explored another approach, in particular, by imposing a variable EoS parameter then finding solutions with different models for $f(Q)$. We have analyzed the solutions with four different $f(Q)$ models and power-law shape function. Our results show that all of these models can provide asymptotically linear EoS. In this view, the first class of solutions ( solutions with linear EoS) can be considered as a special case of solutions with variable EoS when the condition (\ref{38}) is satisfied.
 For some cases of our solutions, we have demonstrated that the energy conditions are violated only in a small region of the space. It has been argued that this region can be in the vicinity of throat or elsewhere. Despite this, we have shown  that  fluid supporting the wormhole can be asymptotically isotropic which seems formidable  for the first class of solutions. In Sec. \ref{sec3}, we have considered a vanishing redshift function, i.e., $\phi(r) = 0$, also solutions with non-constant redshift function are explored in Sec. \ref{sec4}. For the variable redshift function, it was concluded that the complexity of equations does not allow us to construct the wormhole with linear EoS. Besides, many solutions with variable redshift functions lead to a non-physical energy density. Finally, we have explored some solutions with positive energy density which some of them have more consistency in contrast to the others. Solutions with variable redshift function are reachable in the background of $f(Q)$, but the complexity of equations restricts the exploration of these solutions for general forms.

Recently several good papers, which have studied the wormhole in the context of $f(Q)$ gravity, are presented \cite{Wang, Sha, Mus, Hassan, sym, Ban}. The problem of exotic matter is not resolved in these papers in general form; also, solutions with linear EoS are not presented. In this work, we have found solutions with linear EoS, which is highly important in cosmology and wormhole physics. Finding asymptotically flat wormhole solutions with linear EoS is not a simple task in most of the modified theories. As it was mentioned, $f(Q)$ gravity features in second-order field equations; so, it seems that finding asymptotically flat wormhole solutions with linear EoS, in the background of $f(Q)$, is more straightforward in contrast to $f(R)$, $f(R,T)$ and some other modified gravities. Minimizing the violation of energy conditions is the second significant point that is concluded in our paper. While no specific proposal makes a strong case for being highly likely or far better than all others, solutions involving  the minimum violation of energy conditions with a linear EoS may be considered as  the best options until a better alternative comes.  As it was mentioned in the introduction, some of the modified gravities solve the problem of exotic matter i.e., brane cosmology\cite{brane} or Einstein-Cartan gravity \cite{Cartan}, but some others cannot solve this problem entirely yet. For example, some kind of solutions in the context of $f(R)$ gravity or $f(R,T)$ gravity are found which did not respect the energy conditions. On the other hand, the modified gravity should pass many tests to be accepted as an alternative theory for GR. So, we think that discovering the ability of any modified gravity, to describe astrophysical phenomena, is of great importance even though the desired results are not concluded.

To summarize, the $f(Q)$ gravity helps us to study a large class of wormhole solutions that violate  the energy conditions only in some regions of the spacetime. We have utilized a specific $f(Q)$ model,  which can provide analytical exact wormhole solutions with barotropic EoS. By carefully constructing a specific $f(Q)$ function, we have obtained a power-law shape function then using asymptotically flat condition provided the line element of the wormhole.  In addition, we have shown the possibility of existence of  exact asymptotically flat  wormhole solutions, which admits an asymptotically linear EoS, and is isotropic at large scale. This kind of solutions violate all of the energy conditions only in a small region of the spacetime. Since minimizing the  usage of exotic matter for the physical viability of wormhole is  one of the primary goal of studies in wormhole physics, it seems that $f(Q)$ is a good candidate to accomplish this point. Although, wormholes have not been detected yet, in this study, we have  investigated  the possible existence of wormhole geometries in the context of $f(Q)$ gravity.




\begin{thebibliography}{99}

\bibitem{Visser} M. Visser,\textit{ Lorentzian wormholes: From Einstein to Hawking}, (AIP Press, New York, 1995).
\bibitem{lens} R. Shaikh, P. Banerjee, S. Paul, and T. Sarkar, Phys. Lett. B {\bf 789}, 270 (2019); JCAP {\bf 07} 028(2019);  S.N. and N. Riazi Can. J. Phys. {\bf 98}, 1046 (2020); K. Jusufi, A. Övgün, A. Banerjee, and I. Sakallı, Eur. Phys. J. Plus {\bf
    134}, 428 (2019); Farook Rahaman, Ksh. Newton Singh, Rajibul Shaikh, Tuhina Manna, and Somi Akta, Classical Quantum Gravity {\bf 38}, 215007 (2021) .
\bibitem{Rosen} A. Einstein, N. Rosen, Phys. Rev. {\bf48} , 73 (1935).
\bibitem{wheeler}C. W. Misner and J. A. Wheeler, Annals Phys. {\bf2}, 525 (1957).

\bibitem{WH} M. S. Morris, K. S. Thorne, Am. J. Phys. {\bf 56}, 395 (1988).

\bibitem{phantom}R. Lukmanova, A. Khaibullina, R. Izmailov, A. Yanbekov, R.
Karimov, and A. A. Potapov, Indian J. Phys. {\bf 90}, 1319 (2016);
Y. Heydarzade, N. Riazi, and H. Moradpour, Can. J. Phys. {\bf 93}, 1523 (2015);
 F. S. N. Lobo, Phys. Rev. D {\bf 71}, 084011 (2005); S.~V.~Sushkov,  Phys.\ Rev.\ D {\bf 71}, 043520 (2005);O. B. Zaslavskii,  Phys. Rev. D {\bf 72}, 061303(R), (2005);J.A. Gonzalez, F. S. Guzman, N. Montelongo-Garcia, and T.
Zannias, Phys. Rev. D {\bf 79}, 064027 (2009);
P.K. Sahoo, P.H.R.S. Moraes, Parbati Sahoo and  G. Ribeiro, Int. J.  Mod. Phys. D, {\bf 27},1950004 (2018).

\bibitem{phantom1}Francisco S. N. Lobo, Foad Parsaei, and Nematollah Riazi, Phys.\ Rev.\ D {\bf 87}, 084030 (2013).

  \bibitem{cut}M. Visser, S. Kar and N. Dadhich,
Phys. Rev. Lett.  {\bf 90}, 201102 (2003); S. Kar, N. Dadhich and M. Visser, Pramana {\bf 63}, 859 (2004);
E. Eiroa and G. Romero, Gen. Rel. Grav. {\bf 36}, 651 (2004); Phys. Rev. D {\bf 71}, 127501 (2005);Nadiezhda Montelongo Garcia, Francisco S. N. Lobo, and Matt Visser, Phys. Rev. D {\bf 86}, 044026 (2012).
 \bibitem{Remo}Remo Garattini, and Francisco S. N. Lobo, Classical Quantum Gravity {\bf24}, 2401 (2007).
\bibitem{variable} F. Parsaei and S. Rastgoo, Phys. Rev. D {\bf 99}, 104037 (2019).

 \bibitem{foad}F. Parsaei and S. Rastgoo,  Eur. Phys. J. C  {\bf 80}, 366 (2020).
\bibitem{vol}M. Visser, S. Kar and N. Dadhich, Phys. Rev. Lett.  {\bf 90}, 201102 (2003);
S. Kar, N. Dadhich and M. Visser, Pramana {\bf 63}, 859 (2004).
\bibitem{zas}O. B. Zaslavskii, Phys. Rev. D {\bf 76}, 044017 (2007)
\bibitem{rast} H. Moradpour, N. Sadeghnezhad, and S. H. Hendi, Can. J. Phys. {\bf 95}, 1257 (2017).
\bibitem{curvature} N.~M.~Garcia, F.~S.~N.~Lobo,
  Phys.\ Rev.\ D {\bf 82}, 104018 (2010);  Classical Quantum Gravity {\bf 28}, 085018 (2011); Allah Ditta, Ibrar Hussain, G. Mustafa, Abdelghani Errehymy, and Mohammed Daoud, Eur. Phys. J. C  {\bf 81 }, 880 (2021).

  \bibitem{brane}M. L. Camera, Phys. Lett. B {\bf 573}, 27 (2003);
 K. A. Bronnikov and Sung-Won Kim, Phys. Rev. D {\bf67}, 064027 (2003);  F. S. N. Lobo, Phys. Rev. D 75, 064027 (2007); K. C. Wong, T. Harko, and K. S. Cheng, Classical Quantum
Gravity 28, 145023 (2011); Yoshimune Tomikawa, Tetsuya Shiromizu, and Keisuke Izumi
Phys. Rev. D {\bf 90}, 126001 (2014); F. Parsaei, N. Riazi,  Phys.\ Rev.\ D {\bf 91}, 024015 (2015); S. Kar, S. Lahiri,S. SenGupta, Phys. Lett. B {\bf 750}, 319 (2016); F. Parsaei, N. Riazi,  Phys.\ Rev.\ D {\bf 102}, 044003 (2020).

\bibitem{Born}  M. G. Richarte, C. Simeone, Phys. Rev. D {\bf 80}, 104033 (2009); E.F. Eiroa, G.F. Aguirre, Eur. Phys. J. C {\bf 72}, 2240 (2012); Rajibul Shaikh Phys. Rev. D { \bf 98} , 064033 (2018).
\bibitem{quad}Francis Duplessis and Damien A. Easson, Phys. Rev. D {\bf 92}, 043516 (2015).
\bibitem{Cartan}K. A. Bronnikov and A. M. Galiakhmetov, Grav. Cosmol {\bf21}, 283 (2015); Phys. Rev. D {\bf 94}, 124006
(2016) ;  M.R. Mehdizadeh and A.H. Ziaie, Phys. Rev. D {\bf 95}, 064049 (2017).
\bibitem{f(R)}Petar Pavlovic and Marko Sossich, Eur. Phys. J. C  {\bf 75}, 117 (2015);Ernesto F. Eiroa and Griselda Figueroa Aguirre,  Eur. Phys. J. C  {\bf 6}, 132 (2016);
 P.H.R.S. Moraes and P.K. Sahoo,  Phys. Rev. D {\bf  96}, 044038  (2017);
 E. Elizalde and M. Khurshudyan, Phys. Rev. D {\bf  99}, 024051  (2019); Nisha Godani and Gauranga C. Samanta, Eur. Phys. J. C {\bf80}, 30 (2020); F. Parsaei and S. Rastgoo,	arXiv:2110.07278.

\bibitem{Rev}Thomas P. Sotiriou, and Valerio Faraoni, Rev. Mod. Phys. {\bf 82}, 451 (2010).
\bibitem{f(T)}Ksh. Newton Singh, Ayan Banerjee, Farook Rahaman, and M. K. Jasim
Phys. Rev. D {\bf 101}, 084012 (2020).
\bibitem{Jim} J. B. Jimenez, L. Heisenberg, and T. Koivisto, Phys. Rev. D {\bf 98}, 044048 (2018).
\bibitem{Zha}Rui-Hui Lin and Xiang-Hua Zhai, Phys. Rev. D {\bf 103}, 124001 (2021).
\bibitem{Tel}Tiberiu Harko, Tomi S. Koivisto, Francisco S.N. Lobo, Gonzalo J. Olmo, and Diego Rubiera-Garcia,
 Phys. Rev. D {\bf98}, 084043 (2018);Sanjay Mandal, Deng Wang, and P. K. Sahoo
Phys. Rev. D {\bf 102}, 124029 (2020); B. J. Barros et al., Phys. Dark Universe {\bf 30}, 100616
(2020); Noemi Frusciante, Phys. Rev. D {\bf 103}, 044021 (2021).
\bibitem{Ec} Sanjay Mandal, P. K. Sahoo, and J. R. L. Santos, Phys. Rev. D {\bf 102}, 024057 (2020).
\bibitem{Newt} K. Flathmann and M. Hohmann, Phys. Rev. D {\bf 103}, 044030 (2021),


 \bibitem{Black} Fabio D’Ambrosio, Shaun D. B. Fell, Lavinia Heisenberg, and Simon Kuhn
Phys. Rev. D {\bf 105}, 024042 (2022).


\bibitem{Wang}Wenyi Wang, Hua Chen, and Taishi Katsuragawa, Phys. Rev. D {\bf 105}, 024060 (2022).
\bibitem{Sha} Umesh Kumar Sharma, Shweta and Ambuj Kumar Mishra, Int. J of Geo. Methods Mod. Phys.{\bf 02}, 2250019 (2022).
\bibitem{Mus} G.Mustafa, Zinnat Hassan, P.H.R.S.Moraes, P.K.Sahoo, Phys. Lett. B{\bf 821} 136612 (2021).
\bibitem{Hassan}Zinnat Hassan, Sanjay Mandal, and P.K. Sahoo, Fortschr. Phys. {\bf 69}, 2100023 (2021)
\bibitem{sym}Zinnat Hassan, Ghulam Mustafa, and P.K. Sahoo,  Symmetry. {\bf 13}, 1260 (2021).
\bibitem{Ban}Ayan Banerjee, Anirudh Pradhan, Takol Tangphati, Farook Rahaman, Eur. Phys. J. C {\bf 81}, 1031 (2021).









\end{thebibliography}
\end{document}